\documentclass[reqno, 12pt]{amsart}

\usepackage{amsthm,amssymb,amstext,amscd,amsfonts,amsbsy,amsxtra,latexsym, 
amsmath, color}
\usepackage{fullpage}
\usepackage[latin1]{inputenc}%

\definecolor{blue}{rgb}{0,0,1}
\definecolor{red}{rgb}{1,0,0}
\definecolor{green}{rgb}{0,1,0}

\begin{document}

\title{Defeating the Ben-Zvi, Blackburn, and Tsaban Attack on the Algebraic Eraser}

\author{Iris Anshel}
\address{SecureRF Corporation, 100 Beard Sawmill Rd \#350, Shelton, CT 06484}
  \email{ianshel@securerf.com}
  
  \author{Derek Atkins}
\address{SecureRF Corporation, 100 Beard Sawmill Rd \#350, Shelton, CT 06484}
  \email{datkins@securerf.com}

\author{Dorian Goldfeld}
\address{SecureRF Corporation, 100 Beard Sawmill Rd \#350, Shelton, CT 06484 }
\email{dgoldfeld@securerf.com}

\author{Paul E. Gunnells}
\address{SecureRF Corporation, 100 Beard Sawmill Rd \#350, Shelton, CT 06484}
\email{pgunnells@securerf.com}

\date{\today}

\begin{abstract}

\hfill \\
The \emph{Algebraic Eraser Diffie--Hellman} (AEDH) protocol was
introduced in 2005 and published in 2006 by I.~Anshel, M.~Anshel,
D.~Goldfeld, and S.~Lemieux as a protocol suitable for use on
platforms with constrained computational resources, such as FPGAs,
ASICs, and wireless sensors.  It is a group-theoretic cryptographic
protocol that allows two users to construct a shared secret via a
Diffie--Hellman-type scheme over an insecure channel.

Building on the refuted 2012 permutation-based attack of
Kalka--Teichner--Tsaban (KKT), Ben-Zvi, Blackburn, and Tsaban (BBT)
present a heuristic attack, published November 13, 2015, that
attempts to recover the AEDH shared secret.  In their paper BBT
reference the AEDH protocol as presented to ISO for certification (ISO
29167-20) by SecureRF.  The ISO 29167-20 draft contains two profiles
using the Algebraic Eraser.  One profile is unaffected by
this attack; the second profile is subject to their attack
provided the attack runs in real time. This is not the case in most practical deployments.

The BBT attack is simply a targeted attack that does not attempt to
break the method, system parameters, or recover any private
keys. Rather, its limited focus is to recover the shared secret in a
single transaction.  In addition, the BBT attack is based on several
conjectures that are assumed to hold when parameters are chosen
according to standard distributions, which can be mitigated, if not
avoided. This paper shows how to choose special distributions so that
these conjectures do not hold making the BBT attack ineffective for
braid groups with sufficiently many strands. Further, the BBT attack
assumes that certain data is available to an attacker, but there are
realistic deployment scenarios where this is not the case, making
the attack fail completely.  In summary, the BBT attack is flawed
(with respect to the SecureRF ISO draft) and, at a minimum,
over-reaches as to its applicability.

\end{abstract}

\keywords{Algebraic eraser, colored Burau key agreement protocol, group theoretic cryptography, braid groups}

\subjclass{20F36, 94A60} 

\maketitle

\section{Introduction}

The Algebraic Eraser Diffie--Hellman protocol (AEDH, originally AEKAP)
was introduced in 2005 and published in 2006 by Anshel, Anshel,
Goldfeld, and Lemieux \cite{aagl} as a protocol suitable for use on
computationally constrained resource platforms, e.g., FPGAs, ASICs,
and wireless sensors.  More specifically, AEDH is a Group-Theoretic
public-key system designed as a computationally efficient solution for
low-power or passive embedded systems and devices associated with,
among other things, the Internet of Things (IoT).
\vskip 5pt

In November, 2015, Ben-Zvi, Blackburn, and Tsaban (BBT), leveraging
the refuted 2012 permutation-based attack of Kalka, Teichner, and
Tsaban (KTT) \cite{ktt},\footnote{KKT acknowleged the successful
refutation.} published a paper \cite{bbt} claiming a successful
attack against AEDH. In the abstract of \cite{bbt} it is pointed out
that certain implementations of AEDH are proposed as an underlying technology for ISO/IEC
29167-20. The BBT paper presented a heuristic attack that attempts to
recover the AEDH shared secret in a single transaction using
conjectures assumed to hold when parameters are selected based upon a
standard distribution, e.g. a special key choice.  These conjectures
are only asserted by BBT, they give no indication of how they could be proved.  \vskip 5pt

This paper demonstrates that the BBT attack is flawed (with respect to
the SecureRF ISO draft) and, at a minimum, over-reaches as to its
applicability.  The BBT approach fails to run in real time, fails to
take into account that special distributions can be selected to defeat
one of its conjectures, and assumes a certain collection of data is
available to the attacker, which is not always true.
\vskip 5pt

\vskip 5pt
Specifically, the BBT attack is usually
thousands of times slower (on a 4 GHz processor) than the running time
of the AEDH protocol in a constrained device with limited computing
power and memory. The second profile in the ISO 29167-20 draft is an
authentication protocol whereby two users in a lightweight
cryptographic setting run the AEDH key agreement protocol and obtain a
shared secret which is publicly revealed to complete the
authentication. If the BBT attack recovers the shared secret after it
is revealed and authentication is completed, it is of no consequence.
Thus, the attack fails because the information is no longer relevant.
\vskip 5pt

The BBT attack assumes parameters are chosen according to standard
distributions.  Practically, the use of a standard distribution in a
commercial implementation is an over-simplification; realistic
deployment scenarios involve special distributions so that at least
one of the BBT conjectures does not hold.  As a result, the attack
becomes ineffective for braid groups with sufficiently many strands.
\vskip 5pt

The BBT attack assumes that all data required for the attack
is available to an attacker.  As already published in the referenced
ISO specification (29167-20), there exist deployment scenarios in
which this  assumption is false.  Therefore, there are clearly
scenarios that are not subject to the BBT attack.
\vskip 5pt

This paper proceeds as follows.  In Section \ref{s:aedh}, we review
the AEDH protocol.  Next, in Section \ref{s:bbtattack} we summarize
the BBT attack, and in Section \ref{s:defeat} provide two simple bases
for defeating the BBT attack, a conjecture counterexample and
deployment counterexample.  We conclude in Section \ref{s:conclusions}
by noting that the BBT attack is simply a targeted attack that does
not attempt to break the method, system parameters, or recover any
private keys.

\section{The Algebraic Eraser Diffie--Hellman key agreement
protocol.}\label{s:aedh}

Let $B_N$ denote the $N$-strand braid group. Each element in $B_N$
can be expressed as a word in the Artin generators $\{
b_1,b_2,\ldots,b_{N-1} \}$, which are subject to the following
relations: for $i = 1,\ldots,N - 1$, we have
\begin{equation}\label{eq:rel1}
b_i b_{ i+1 } b_i  =  b_{i+1} b_i b_{i+1},
\end{equation}
\noindent
and for all $i,j $  with  $ |i-j| \ge 2$, we have  
\noindent
\begin{equation}\label{eq:rel2}
b_i b_j  =  b_j b_i.
\end{equation}
Thus any $\beta \in B_{N}$ can be expressed as a product of the form  
\begin{equation}\label{eq:artinexpression}
\beta = b_{ i_1}^{\epsilon_1} \; b_{i_2 }^{\epsilon_2}\; \cdots\;
b_{i_k }^{\epsilon_k },
\end{equation}
where $i_j \in \{1,\ldots,N-1\}$, and
$\epsilon_j \in \{\pm 1 \}$.

Each braid $\beta \in B_{N}$ determines a permutation in $S_{N}$, the
group of permutations of $N$ letters, as follows.  Let $\sigma_i \in
S_N$ be the $i$th simple transposition, which maps $i \mapsto i+1,
i+1\mapsto i$, and leaves $\{1,\ldots,i-1,i+2,\ldots,N\}$ fixed.  Then
if $\beta \in B_N$ is expressed as in (\ref{eq:artinexpression}), we
map $\beta$ to the permutation $\sigma_\beta = \sigma_{i_1}\dotsb
\sigma_{i_k}$.

Fix a prime power $q$, let $F_{q}$ be the finite field of order $q$,
and let $t_{1},\ldots ,t_{N}$ be indeterminates.  Consider the group
of invertible $N\times N$ matrices with entries in the field of
rational functions $F_{q} (t_{1},\dotsc ,t_{N})$, denoted ${\mathcal
M}$. Observe the permutation group $S_N$ acts on $\mathcal{M}$ by
permuting the variables, and we can hence form the semidirect product
${\mathcal M} \rtimes S_N $. For $i=1,\dotsc , N-1$, let $CB (b_{i})$
be the matrix defined by
\begin{equation}\label{eq:cbmatrix}
CB (b_{i}) = \left(\begin{array}{ccccc}
1&&&&\\
&\ddots&&&\\
&t_{i}&-t_{i}&1&\\
&&&\ddots&\\
&&&&1
\end{array} \right),
\end{equation}
where the indicated variables appear in row $i$, and if $i=1$ the leftmost $t_{1}$
is omitted.  We similarly define $CB (b_{i}^{-1})$ by modifying \eqref{eq:cbmatrix} slightly:
\[
CB (b_{i}^{-1}) = \left(\begin{array}{ccccc}
1&&&&\\
&\ddots&&&\\
&1&-t_{i+1}^{-1}& {t_{i+1}^{-1}}&\\
&&&\ddots&\\
&&&&1
\end{array} \right).
\]
With these matrices in place, each braid generator $b_i$ (respectively, inverse generator
$b_{i}^{-1}$) determines a Colored Burau/permutation pair
$(CB(b_i),\sigma_i )$ (resp., $(CB (b_{i}^{-1}), \sigma_{i})$), and we
consider the subgroup of  ${\mathcal M} \rtimes S_N $ generated by
these colored Burau pairs. Since the pairs $(CB(b_i),\sigma_i )$
satisfy the braid relations \eqref{eq:rel1}--\eqref{eq:rel2}, the natural mapping  
$$b_i \longmapsto (CB(b_i),\sigma_i )$$
defines a representation $B_N \rightarrow {\mathcal M} \rtimes S_N$,
called the Colored Burau representation.  By fixing a collection of nonzero elements, termed t-values, 
$$\{ \tau_1,\tau_2,\ldots,\tau_N \}\subset F_q,$$ the mapping $t_i
\mapsto \tau_i$ induces a homomorphism $$\Pi\colon {\mathcal M}
\rightarrow GL_N(F_{q}).$$ The operation {\it e-multiplication} is the
right action of the group of colored Burau pairs on the direct product
$GL_N(F_{q}) \times S_N$ and is defined by
$$(m, \sigma_0) \star (CB(\beta), \sigma_{\beta}) \;=\; (m\cdot \Pi(\phantom{}^{\sigma_0} \beta), \sigma_0 \sigma_{\beta }).$$

 With these definitions in place, users Alice and Bob execute the  AEDH protocol as follows:
 \vskip 5pt\noindent
 {\bf System Data:}
 \vskip 5pt
\begin{itemize}
\setlength\itemsep{1em}
\item A public matrix $m_0\in GL_N(F_{q})$.
\item A set of  public t-values $\{\tau_1, \tau_2, \ldots,\tau_N\} \subset   F_q^\times$.
\item Two sets of user conjugates in $B_N$, one of which is public: 
$$\{z a_1z^{-1}, z a_2z^{-1}, \ldots, z a_kz^{-1}\} \quad \text{and} \quad 
 \{z b_1z^{-1}, z b_2z^{-1}, \ldots, z b_{\ell}z^{-1}\}.$$ 
 It is assumed that the elements $a_i,b_j$ commute, and that the
 conjugates are suitably rewritten so that the element  $z \in B_N$
 remains unknown to the users.
\end{itemize}

\vskip 5pt\noindent
 {\bf Private/Public Keys:}
 \vskip 5pt
\begin{itemize}
\setlength\itemsep{1em}
\item  Each user chooses a random private matrix $m_A, m_B$ of the form
$$m_A \;=\; \sum f_i m_0^i , \quad m_B \;=\; \sum g_i m_0^i \in F_q[m_0].$$  
\item 
Each user chooses a random private braid word $w_A$, $w_B$ in their respective set of conjugates:
\begin{align*}
w_A &\;\in\; \langle z a_1z^{-1}, z a_2z^{-1}, \ldots, z a_kz^{-1}\rangle,\\
w_B &\;\in\; \langle z b_1z^{-1}, z b_2z^{-1}, \ldots, z b_{\ell}z^{-1}\rangle.
\end{align*}
\item The user private keys are given by $(m_A, w_A), (m_B, w_B).$ It is assumed that the user whose conjugates are public will use ephemeral private keys.
\item Each user produces and exchanges their respective public  keys:
\begin{align*}{\text {Pub}}_A &\;=\; (m_A, 1) \star (CB(w_A), \sigma_{w_A}),\\
{\text {Pub}}_B &\;=\; (m_B, 1) \star (CB(w_B), \sigma_{w_B}).\end{align*}
\item Both users evaluate the shared secret $K$ using the received public keys and  their private keys:
$$K = (m_A,1)\cdot Pub_B \star ( CB(w_A), \sigma_{w_A}  ) = (m_B,1)\cdot  Pub_A\star ( CB(w_B), \sigma_{w_B}).$$
\end{itemize}

\section{The Ben-Zvi--Blackburn--Tsaban Attack}\label{s:bbtattack}

The BBT attack proceeds as follows.  The
attacker Eve requires access to the public keys $ \text{Pub}_{A},  \text{Pub}_{B}$ of
Alice and Bob and any other information transmitted over the insecure
channel.  This includes the  t-values $\tau_{1},\ldots ,\tau_{N}$, the
matrix $m_{0}$, and Alice's conjugates $\{za_{i}z^{-1} \mid i = 1, \ldots, k\}$.  
Let $A$ be
the subgroup of $B_{N}$ generated by Alice's conjugates.  Let $C$ be
the subspace $ F_{q}[m_{0}] \subset M_{N} ( F_{q})$ of all polynomials
in $m_{0}$ over $ F_{q}$, where $M_{N} ( F_{q})$ denotes the vector
space of $N\times N$ matrices over $ F_{q}$.

Let $P\subset A$ be the subgroup of {\it pure braids} in Alice's
subgroup.  In particular, $P$ is the kernel of the
natural projection $B_{N}\rightarrow S_N$ is restricted to $A$.  The group $P$
determines a subspace $V \subset M_{N} (F_{q})$: we take the subspace
spanned by the images $\Pi (w)$ as $w$ ranges over $P$.

Let $ \text{Pub}_{A} = (p,g)$ be Alice's public key, and let $ \text{Pub}_{B} =
(q,h)$ be Bob's public key.  To recover the shared secret $K$, Eve
plans to find 
\begin{itemize}
\setlength\itemsep{1em}
\item a matrix $\tilde{c} \in C$,
\item an element $\alpha '\in V$, and 
\item an element $(\tilde{a}, g) \;\in\; \langle (CB(\beta),\sigma_{\beta})\;|\; \beta \in B_N \rangle$

\end{itemize}
\noindent
satisfying the following property:
\[
(p,g) = \tilde{c}\cdot (\alpha ',1)\star (\tilde{a},g).
\]
In other words, Eve seeks a factorization of Alice's public key.
Using this data, she then takes pure braids $\alpha_{i} \in P$ such
that the elements $\Pi (\alpha_{i})$ give a basis of $V$, and writes $\alpha'$ as a linear
combination 
\[
\alpha ' = \sum_{i} \lambda_{i} \Pi (\alpha_{i}), \quad \lambda_{i}\in F_{q}.
\]
She then forms the matrix 
\[
\beta ' = \sum_{i} \lambda_{i} \Pi (\phantom{}^{h}\alpha_{i}),
\]
where the superscript denotes the action of $h$ in the semidirect
product.  Then  BBT prove that the shared secret $K$ can be expressed
as 
\[
K = \tilde{c}\cdot (q\beta ', h) \star (\tilde{a},g).
\]

As an example, BBT applied their attack to
sample data on the braid group $B_{16}$ and with the finite field
$ F_{256}$.  The authors successfully recovered the shared secret
after a running time of approximately $8$ hours and using less than 64
MB of memory.

Eve finds the elements $\tilde{c}$, $\alpha '$, and
$\tilde{a}$ needed above to reconstruct $K$ in the following way:
\vskip 5pt
\begin{itemize}
\setlength\itemsep{1em}
\item \textbf{Precomputation stage.} Eve first determines a basis of
$V$.  Let $\mu_{i} = (z\alpha_{i}z^{-1}, g_{i})$ be the elements of
${\mathcal M} \rtimes S_N$ corresponding to Alice's conjugates, which are known.  Eve requires a method   to produce elements $g$ in $S_{N}$ that have order
$r\leq N$ and that are short products of the $g_{i}$. BBT assumes this is always possible. Write $g = \prod g
(j)^{\varepsilon_{j}}$, where each $g (j)\in \{g_{1},\dotsc,g_{k} \}$
and each $\varepsilon_{j}\in \{\pm 1 \}$, and let $\mu (j)$ be the
$\mu_{i}$ corresponding to $g (j)$.  Then since $g^{r}$ is trivial,
the product $\alpha = (\prod \mu (j)^{\varepsilon_{j}})^{r}$ is pure.
Eve constructs many matrices of the form $\Pi (\alpha)$ and stops when
she has a set $\alpha_{1},\dotsc ,\alpha_{m}$ such that the dimension
of the span of the $\Pi (\alpha_{i})$ has stabilized.\vskip5pt

\item \textbf{Stage 1: Finding $\tilde{a }$.}  Let $(p,g)$ be Alice's
public key.  Eve again must find a
product of the $\mu_{i}$ of the form $(\tilde{a},g)$, that is a
product with second component equal to $g$.  This is done using the
algorithms in \cite{ktt}.
\vskip5pt
\item \textbf{Stage 2: Finding $\tilde{c}$.}  Define a matrix $\gamma$
by $(\gamma ,1) = (p,g)\star (\tilde{a},g)^{-1}$.  The matrix
$\tilde{c}$ is then taken to be an invertible element of the
intersection $C\cap \gamma V$.\vskip5pt

\item \textbf{Stage 3: Remaining parameters.}  The matrix
${\alpha'}$ is defined to be $\tilde{c}^{-1}\gamma$.  
\end{itemize}

\section{Defeating the Ben-Zvi--Blackburn--Tsaban Attack}\label{s:defeat}

\subsection{Conjecture Counterexample} 
\hfill \\

\smallskip
As detailed above, in the BBT attack it is assumed that there are many
short expressions in the publicly known Reader conjugates which are
associated with low order permutations. While this statement holds
most of the time, it is not correct in every instance.  At the end of
their paper, BBT indicate that it may be possible to immunize the
Algebraic Eraser against their attack by working with very carefully
chosen presentations.  The following demonstrates how to effectively
produce a set of permutations whose short expressions have high order
most of the time, countering their conjecture. The values of the
parameter $N$ will necessarily be in a higher range than is discussed
in BBT, but the method can be computationally viable by applying
suitable projection operators associated to singular first private
matrices.  

\vskip 10pt To counter the conjecture produce a set of
$k\ge 2$ permutations $$\rho_1, \rho_2, \ldots,\rho_k$$ in the
symmetric group $S_N$ with the property that almost all short
expressions (short words in $\rho_1, \rho_2, \ldots,\rho_k$) have
order greater than $$ \alpha_N\cdot e^{\frac12\cdot \sqrt{N \log
N}},$$ for some constant $\alpha_N > 0.$ The constant $\alpha_N$ is
yet to be determined, but initial testing shows it is not too small.

\vskip 10pt Let $p_N$ denote the largest prime such that the sum of
the primes less than $p_N$ does not exceed $N$. By the prime number
theorem, we have $p_N \sim \sqrt{N\log N}.$ \vskip 5pt We then
define \begin{align*}
  \rho_1 & := c_1(3)\, c_1(5)\, c_1(7) \cdots c_1(p_N),\\
  \rho_2 & := c_2(3)\, c_2(5)\, c_2(7) \cdots c_2(p_N),\\
 &\phantom{x.} \vdots\\
  \rho_k & := c_k(3)\, c_k(5)\, c_k(7) \cdots c_k(p_N).
  \end{align*}
  \vskip 3pt
   Here, for every prime $p$ and any $1\le i\le k$, we let $c_i(p)$ denote a $p$-cycle in $S_N$. The $p$-cycles $c_i(p)$ can be randomly chosen and are assumed to satisfy the following additional properties:
 
\begin{itemize}
\setlength\itemsep{1em}
{\it
\item For each $1\le i\le k$, the cycles $c_i(3),c_i(5),\ldots, c_i(p_N)$ are disjoint.
   
\item For each prime $3\le p\le p_n$, the cycles $c_1(p), c_2(p), \ldots, c_k(p)$ all have the same fixed points.
   
\item For each prime $3\le p\le p_n$, no two of the cycles $c_1(p),
   c_2(p), \ldots, c_k(p)$ are integer powers of each other.}
\end{itemize}
   
   \vskip 10pt
   Since the cycles are disjoint, the order of each $\rho_i\, (1\le i\le k)$ is given by
   the product of primes $$\prod_{3\le p \le p_N} p \; \sim \; e^{p_N} \; \sim \;\frac12\cdot e^{\sqrt{N\log N}},$$
   where the above asymptotic formula again follows from the Prime Number Theorem.
     
   \vskip 10pt
 
To facilitate working with large braid groups one can choose a highly singular seed matrix $m_0$ in AEDH which has the property that for any $g \in GL_{N}(F_q)$, the matrix product
$$m_0 \cdot g$$ projects onto a submatrix of $g$ consisting of $r$ rows of $g$ for some small $1 < r<N.$ This reduces all the public keys
to matrices of size $r \times N$ instead of $N^2.$ In addition, it is possible to work with a much smaller finite field when $N$ is large, and, hence, arrive at manageable public key sizes.   It remains to determine optimal parameters for specific applications. This will be the topic of a future paper.

\subsection{Deployment Counterexample} \hfill {\it } \\

The BBT attack requires knowledge of both public keys, the t-values,
the seed matrix $m_0$, and one set of conjugates.  Lack of any single
one of these items will defeat the attack.
\vskip 5pt

The BBT paper references the ISO 29167-20 draft specification of the
Algebraic Eraser.  That specification contains two deployment profiles
for AEDH.  In one of the profiles one party has access to a database
that contains public key material for the other parties. Specifically,
in this profile an attacker never has access to one of the public keys
and, as a result, cannot mount the attack to derive the shared secret.
Other deployment scenarios also exist where an attacker does not have
access to one or more pieces of data required to mount the attack.  In
all these scenarios the attack cannot succeed.
 
\section{Conclusions}\label{s:conclusions}

The BBT attack is simply a targeted attack that does not attempt to
break the method, system parameters, or recover any private keys.  Its
limited focus attempts to recover the shared secret in a single transaction for a
class of weak keys.  The attack is based on several conjectures, none
of which are proven.  As per one BBT conjecture, when conjugate material
is chosen poorly the attacker can find short expressions and mount an
attack.  However, when the conjugates are chosen with specific classes
of permutations the conjecture fails as does the BBT attack for braid
groups with sufficiently many strands.  Finding counterexamples of
other BBT conjectures is left for a future paper.
\vskip 5pt

Similarly, deployment scenarios (such as one of the profiles in ISO
29167-20) deprive an attacker from the information required to mount
the attack.  Without all required data the attack cannot even begin.
\vskip 5pt

Therefore, AEDH, a group-theoretic cryptographic protocol that
constructs a shared secret via a Diffie--Hellman-type scheme, is 
secure for many practical applications, including platforms with
constrained computation resources, such as FPGAs, ASICs, and wireless
sensors.

\bibliographystyle{amsplain_initials} \bibliography{responsetoBBT}

% \begin{thebibliography}{3}
   
%  \bibitem{AAGL} Anshel, Iris;  Anshel, Michael;   Goldfeld, Dorian; and  Lemieux, Stephane, {\it Key agreement, the Algebraic Eraser$^{TM}$, and Lightweight Cryptography,}  Algebraic methods in cryptography, Contemp. Math., vol. 418, Amer. Math. Soc., Providence, RI, 2006, pp. 1--34.
 
%   \vskip 5pt
  
%    \bibitem{AEHash} Anshel, Iris;   Goldfeld, Dorian, {\it Cryptographic hash function,} US Patent number 8,972,715, March 3, 2015.
   
%    \vskip 5pt
  
%  \bibitem{BKL} Birman, Joan; Ko, Ki Hyoung; Lee, Sang Jin; {\it A new approach to the word and conjugacy problems in the braid groups,} Adv. Math. 139 (1998), no. 2, 322--353.

% \vskip 5pt \bibitem{BBT} A.~Ben-Zvi, S.~Blackburn, and B.~Tsaban,
% {\it A Practical Cryptanalysis of the {Algebraic Eraser},} preprint,
% November, 2015.
% \vskip 5pt
% \bibitem{KTT}
% A.~Kalka, M.~Teicher, and B.~Tsaban, \emph{Short expressions of permutations as
%   products and cryptanalysis of the {A}lgebraic {E}raser}, Adv. in Appl. Math.
%   \textbf{49} (2012), no.~1, 57--76.

%    \end{thebibliography}
  
\end{document}